\documentstyle[preprint,aps,prd]{revtex}
\tightenlines
\def\be{\begin{equation}}
\def\ee{\end{equation}}
\def\ba{\begin{eqnarray}}
\def\ea{\end{eqnarray}}
\begin{document}
\draft 
\title{Riemannian and Teleparallel Descriptions of the Scalar Field
Gravitational Interaction}
\author{V. C. de Andrade and J. G. Pereira}
\vskip 0.5cm
\address{Instituto de F\'{\i}sica Te\'orica\\
Universidade Estadual Paulista\\
Rua Pamplona 145\\
01405-900\, S\~ao Paulo \\ 
Brazil}
\maketitle

\begin{abstract}
A comparative study between the metric and the teleparallel
descriptions of gravitation is made for the case of a scalar field.
In contrast to the current belief that only spin matter
could detect the teleparallel geometry, scalar matter
being able to feel the metric geometry only, we show
that a scalar field is able not only to feel anyone of these geometries, but
also to produce torsion. Furthermore, both descriptions are found to be
completely equivalent, which means that in fact, besides coupling to
curvature, a scalar field couples also to torsion.
\end{abstract}

\section{Introduction}

Since the early days of general relativity, the description of the
gravitational interaction has been deeply connected to the geometry 
of spacetime. According to its postulates, the presence of gravitation
produces a curvature in spacetime, the gravitational interaction being
achieved by supposing a particle to freely follow its geodesics. Curvature,
therefore, is considered to be an intrinsic attribute of spacetime.

On the other hand, theoretical developments have since long evoked the
possibility of including torsion in the description of the gravitational
interaction. In the usual approach to gravitation, torsion is set to vanish
from the very beginning, and there seems to be no compelling experimental
evidence not to set this condition. However, as we are going to see, in the
context of the teleparallel equivalent of general relativity~\cite{hayshi},
one can not set the vanishing of torsion without vanishing the curvature as
they are manifestations of the same gravitational field. In other words, the
vanishing of torsion only, would spoil the alluded equivalence.

Curvature and torsion present completely different characteristics from
the point of view of the gravitational interaction. Curvature, according
to general relativity, is used to {\it geometrize} spacetime, and in this
way successfully describe the gravitational interaction. On the other hand,
teleparallelism attributes gravitation to torsion, but in this case torsion
accounts for gravitation not by geometrizing the interaction, but by
acting as a force~\cite{paper}. This means that, in the  teleparallel
equivalent of general relativity, there are no  geodesics, but force
equations quite analogous to the Lorentz force equation of electrodynamics.

Differently from what is usually done in general relativity, in what
follows we will benefit by separating the notions of space and
connections. From a formal point of view, curvature and torsion are in
fact properties of a connection~\cite{koba}, and a great many connections
may be defined on the same space~\cite{livro}. Strictly speaking, there is
no such a thing as curvature or torsion of spacetime, but only curvature or
torsion of connections. This becomes evident if we notice that different
particles feel different connections, and consequently show distinct
trajectories in spacetime. In the general relativity case, though, there
is a point for taking the Levi-Civita connection of spacetime as
part of its definition: universality of gravitation implies that all
particles feel it the same, and this makes possible to interpret the
curvature of the  connection as the curvature of spacetime itself, leading
thus to the general relativity scene. It seems far wiser, however, to take
spacetime simply as a manifold, and connections (with their curvatures and
torsions) as additional structures.

With the purpose of exploring the interaction of gravitation with a 
scalar field, as well as the role played by curvature and torsion in the
description of this interaction, we assume the background of this work to be
a spacetime manifold on which a nontrivial tetrad field is defined. The
context may be, for example, that of a gauge theory for the translation
group~\cite{paper}, which is the most common prototype of a tetrad theory.
In this context, the gravitational field appears as the nontrivial part of
the tetrad field. We will use the greek alphabet ($\mu$, $\nu$,
$\rho$,~$\cdots=1,2,3,4$) to denote tensor indices,  that is, indices
related to spacetime. The latin alphabet ($a$, $b$, $c$,~$\cdots=1,2,3,4$)
will be  used to denote local Lorentz (or tangent space) indices. Of course,
being of the same kind, tensor and local Lorentz indices can be changed
into each other with the use of the tetrad, denoted by $h^{a} {}_{\mu}$,
and supposed to satisfy
\be
h^{a}{}_{\mu} \; h_{a}{}^{\nu} = \delta_{\mu}{}^{\nu} \quad
; \quad h^{a}{}_{\mu} \; h_{b}{}^{\mu} =
\delta^{a}{}_{b} \; .
\label{orto}
\ee

\section{The Spacetime Geometry}

As already discussed, curvature and torsion are properties of a connection,
and many  different connections may be defined on the same space. For
example, denoting by $\eta_{a b}$ the metric tensor of the tangent space, a
nontrivial tetrad field can be used to define the riemannian metric
\be
g_{\mu \nu} = \eta_{a b} \; h^a{}_\mu \; h^b{}_\nu \; ,
\label{gmn}
\ee
in terms of which the Levi--Civita connection 
\be
{\stackrel{\circ}{\Gamma}}{}^{\sigma}{}_{\mu \nu} = \frac{1}{2} 
g^{\sigma \rho} \left[ \partial_{\mu} g_{\rho \nu} + \partial_{\nu}
g_{\rho \mu} - \partial_{\rho} g_{\mu \nu} \right]
\label{lci}
\ee
can be introduced. Its curvature
\be
{\stackrel{\circ}{R}}{}^{\theta}{}_{\rho \mu \nu} = \partial_\mu
{\stackrel{\circ}{\Gamma}}{}^{\theta}{}_{\rho \nu} +
{\stackrel{\circ}{\Gamma}}{}^{\theta}{}_{\sigma \mu}
\; {\stackrel{\circ}{\Gamma}}{}^{\sigma}{}_{\rho \nu} - (\mu
\leftrightarrow \nu) \; ,
\label{rbola}
\ee
according to general relativity, accounts exactly for the gravitational 
interaction. Owing to the universality of gravitation, which means that
all particles feel ${\stackrel{\circ}{\Gamma}}{}^{\sigma}{}_{\mu \nu}$ the
same, it turns out possible to describe the gravitational interaction by
considering a Riemann spacetime with the curvature of the Levi--Civita
connection, in which all particles will follow geodesics. This is the stage
set of Einstein's General Relativity, the gravitational interaction being
mimicked by a geometrization of spacetime. 

On the other hand, a nontrivial tetrad field can also be used to define the
linear Cartan connection 
\be
\Gamma^{\sigma}{}_{\mu \nu} = h_a{}^\sigma \partial_\nu
h^a{}_\mu \; ,
\label{car}
\ee
with respect to which the tetrad is parallel:
\be
{\nabla}_\nu \; h^{a}{}_{\mu} \equiv
\partial_\nu h^{a}{}_{\mu} - \Gamma^{\rho}{}_{\mu \nu} \,
h^{a}{}_{\rho} = 0 \; .
\label{weitz}
\ee
For this reason, tetrad theories have received the name of
teleparallelism, or absolute parallelism. Plugging in Eqs.(\ref{gmn}) and
(\ref{lci}), we get
\be
{\Gamma}^{\sigma}{}_{\mu \nu} = {\stackrel{\circ}{\Gamma}}{}^
{\sigma}{}_{\mu \nu} + {K}^{\sigma}{}_{\mu \nu} \; ,
\label{rel}
\ee
where
\be
{K}^{\sigma}{}_{\mu \nu} = \frac{1}{2} \left[
T_{\mu}{}^{\sigma}{}_{\nu} + T_{\nu}{}^{\sigma}{}_{\mu} -
T^{\sigma}{}_{\mu \nu} \right]
\label{conto}
\ee
is the contorsion tensor, with
\be
T^\sigma{}_{\mu \nu} = \Gamma^{\sigma}{}_{\nu \mu} - \Gamma^
{\sigma}{}_{\mu \nu} \; 
\label{tor}
\ee
the torsion of the Cartan connection. If now, analogously to the  way the
Riemann spacetime was introduced, we try to introduce a spacetime with the
same properties of the Cartan connection $\Gamma^{\sigma}{}_{\nu \mu}$, we
end up with a Weitzenb\"ock spacetime~\cite{weitz}, a space presenting
torsion, but no curvature. This spacetime is the stage set of the
teleparallel description of gravitation. Considering that local
Lorentz indices are raised and lowered with the Minkowski metric $\eta^{a
b}$, tensor indices on it will be raised and lowered with  the riemannian
metric $g_{\mu \nu} = \eta_{a b} \; h^a{}_\mu \; h^b{}_\nu$~\cite{hayshi}.
Universality of gravitation, in this case, means that all particles feel
$\Gamma^{\sigma}{}_{\nu \mu}$ the same, which in turn means that they will
also feel torsion the same. 

{}From the above considerations, we can infer that the presence of a
nontrivial tetrad field induces both, a riemannian and a teleparallel
structures in spacetime. The first is related to the Levi--Civita
connection, a connection presenting curvature, but no torsion. The second
is related to the Cartan connection, a connection presenting torsion, but
no curvature. It is important to remark that both connections are defined
on the very same spacetime, a spacetime endowed with both a riemannian and
a teleparallel structures.  

As already remarked, the curvature of the Cartan connection vanishes
identically:
\be
{R}^{\theta}{}_{\rho \mu \nu} = \partial_\mu
{\Gamma}^{\theta}{}_{\rho \nu} + {\Gamma}^{\theta}{}_{\sigma
\mu} \; {\Gamma}^{\sigma}{}_{\rho \nu} - (\mu  \leftrightarrow \nu)
\equiv 0 \; .
\label{r}
\ee
Substituting ${\Gamma}^{\theta}{}_{\mu \nu}$ from
Eq.(\ref{rel}), we get
\be
{R}^{\theta}{}_{\rho \mu \nu} =
{\stackrel{\circ}{R}}{}^{\theta}{}_{\rho \mu \nu} +
Q^{\theta}{}_{\rho \mu \nu} \equiv 0 \; ,
\label{eq7}
\ee
where ${\stackrel{\circ}{R}}{}^{\theta}{}_{\rho \mu \nu}$ is the curvature
of the Levi--Civita connection, and
\be
Q^{\theta}{}_{\rho \mu \nu} = D_\mu K^{\theta}{}_{\rho \nu} +
{K}^{\theta}{}_{\sigma \mu} \; K^{\sigma}{}_{\rho \nu} -
(\mu  \leftrightarrow \nu)
\ee
with
\be
D_\mu K^{\theta}{}_{\rho \nu} =
\partial_\mu {K}^{\theta}{}_{\rho \nu} +
\left({\Gamma}{}^{\theta}{}_{\sigma \mu} - {K}^{\theta}{}_{\sigma \mu}
\right) K^{\sigma}{}_{\rho \nu} -
\left({\Gamma}{}^{\sigma}{}_{\rho \mu} - {K}^{\sigma}{}_{\rho \mu}
\right) K^{\theta}{}_{\sigma \nu} \; .
\ee
Equation (\ref{eq7}) has an interesting interpretation: the
contribution ${\stackrel{\circ}{R}}{}^{\theta}{}_{\rho \mu
\nu}$ coming from the Levi--Civita connection, compensates
exactly the contribution $Q^{\theta}{}_{\rho \mu \nu}$ coming from the
contorsion tensor, yielding an identically zero Cartan curvature tensor
${R}^{\theta}{}_{\rho \mu \nu}$. This is a constraint satisfied by the
Levi--Civita and Cartan connections, and is the fulcrum of the equivalence
between the riemannian and the teleparallel descriptions of gravitation.

\section{General Relativity and its Telepar\-al\-lel \\
E\-quiv\-a\-lent}

According to general relativity, the description of the interaction
between scalar matter and gravitation requires a spacetime endowed with a
riemannian structure. The dynamics of the gravitational field, in this
case, turns out to be described by a variational principle with the
lagrangian
\be
{\cal L}_G = \frac{\sqrt{-g} \, c^4}{16 \pi G} \; {\stackrel{\circ}{R}} \; ,
\label{ehl}
\ee
where ${\stackrel{\circ}{R}}= g^{\mu \nu}
{\stackrel{\circ}{R}}{}^{\rho}{}_{\mu \rho \nu}$ is the scalar curvature of
the Levi--Civita connection, and $g={\det}(g_{\mu \nu})$. This lagrangian,
which depends on the Levi-Civita connection only, can be rewritten in an
alternative form depending on the Cartan connection only. In fact,
substituting
${\stackrel{\circ}{R}}$ as obtained from Eq.(\ref{eq7}), up to
divergences we obtain
\be
{\cal L}_G = \frac{h c^4}{16 \pi G} \; \left[\frac{1}{4} \;
T^\rho{}_{\mu \nu} \; T_\rho{}^{\mu \nu} + \frac{1}{2} \;
T^\rho{}_{\mu \nu} \; T^{\nu \mu} {}_\rho - T_{\rho \mu}{}^{\rho}
\; T^{\nu \mu}{}_\nu \right] \; ,
\label{lagr3}
\ee
where $h={\det}(h^a{}_\mu)=\sqrt{-g}$. This is exactly the lagrangian
of a gauge theory for the translation group~\cite{paper}, which means
that a translational gauge theory, with a lagrangian quadratic in the
torsion field, is completely equivalent to general relativity, with its
usual lagrangian linear in the  scalar curvature~\cite{maluf}. As a
consequence of this equivalence, therefore, gravitation might have two
equivalent descriptions, one in terms of a metric geometry, and another
one in which the underlying geometry is that provided by a teleparallel
structure. It is important to remark that, in this approach, the
lagrangian (\ref{lagr3}) has been obtained without requiring it to be
local Lorentz invariant. The usual criticism~\cite{ha} about the deduction
of that lagrangian~\cite{cho}, therefore, does not apply here.

In the absence of gravitational field, the tetrad becomes trivial, $g_{\mu
\nu}$ becomes the Minkowski metric, and both the curvature 
${\stackrel{\circ}{R}}{}^{\theta}{}_{\rho \mu \nu}$ as well as the torsion
$T^\theta{}_{\mu \nu}$ vanish simultaneously. In other words, it is not
possible to set a vanishing torsion without having a vanishing curvature, as
they are manifestations of the same gravitational field. It is important to
remember at this point that curvature and torsion are geometrical properties
of different connections. There is no a connection presenting simultaneously
non--vanishing curvature and torsion, which means that no Riemann--Cartan
spacetime enters the description of the gravitational interaction.
Furthermore, according to our approach, we can say that general relativity
does not assume a vanishing torsion: despite always present, it simply does
not make use of it. On the other hand, in consonance to what happens with
the lagrangian of the gravitational field, there exists an alternative
description of gravitation, the so called teleparallel description, which
makes use of torsion only, not curvature. In this sense, the Cartan
connection can be considered as a kind of {\it dual} to the Levi--Civita
connection, the riemannian--teleparallel equivalence being a kind of
{\it dual symmetry} presented by gravitation.

\section{Scalar Fields: Laplace--Beltrami and its \\
Teleparallel Equivalent}

Let us consider the lagrangian for a free scalar field $\phi$ in a Minkowski
spacetime~\cite{birrel}:
\be
{\cal L}_{\phi} = \frac{1}{2} \left[ \eta^{a b} \; \partial_a \phi
\; \partial_b \phi - m^2 \phi^2 \right] \; .
\label{lfree}
\ee
According to the usual minimal coupling prescription, which brings the
free lagrangian to a lagrangian written in terms of the riemannian structure
of spacetime, the gravitational interaction can be obtained through the
replacements
\begin{eqnarray}
\eta^{a b} \; &\longrightarrow& \; g^{\mu \nu}  \\
{} \nonumber \\
\partial_a  \; &\longrightarrow& \; {\stackrel{\circ}{\nabla}}_{\mu} \; , 
\end{eqnarray}
with $g^{\mu \nu}$ a riemannian metric, and ${\stackrel{\circ}
{\nabla}}_{\mu}$ the Levi-Civita covariant derivative which, for the
specific case of a scalar field, is simply an ordinary derivative. Therefore,
in terms of the riemannian structure, the lagrangian describing a scalar
field in interaction with gravitation turns out to be
\be
{\cal L}_{\phi} = \frac{\sqrt{- g}}{2} \left[ g^{\mu \nu} \, \partial_\mu \phi
\; \partial_\nu \phi - m^2 \phi^2 \right] \; .
\label{lriem}
\ee
By using the identity
$$
\partial_\mu \sqrt{- g} = \frac{\sqrt{- g}}{2} \, g^{\rho \lambda} \partial_\mu
g_{\rho \lambda} \equiv \sqrt{- g} \; {\stackrel{\circ}
{\Gamma}}{}^{\rho }{}_{\mu \rho} \; ,
$$
it is easy to show that the corresponding field equation is
\be
{\stackrel{\circ}{\Box}} \phi + m^2 \phi = 0 \; ,
\label{klein0}
\ee
where
\be
{\stackrel{\circ}{\Box}} \phi = {\stackrel{\circ}{\nabla}}_{\mu} \,
\partial^\mu \phi \equiv \frac{1}{\sqrt{- g}} \; \partial_\mu \left(
{\sqrt{- g}} \; g^{\rho \mu} \; \partial_\rho \phi \right)
\label{labe}
\ee
is the Laplace--Beltrami derivative of $\phi$, with
\be
{\stackrel{\circ}{\nabla}}_{\mu} = \partial_\mu + {\stackrel{\circ}
{\Gamma}}{}^{\rho }{}_{\mu \rho}
\ee
the expression for the Levi--Civita covariant divergence of
$\partial^\mu \phi$. We notice in passing that it is completely equivalent
to apply the minimal coupling prescription in the lagrangian or in the
field equations. In a locally inertial coordinate system, the Levi--Civita
connection vanishes, and the Laplace--Beltrami becomes the free--field
d'Alambertian operator. This is the usual version of the (strong)
equivalence principle~\cite{weinberg}.

Let us consider now the total lagrangian
${\cal L}={\cal L}_G + {\cal L}_{\phi}$, with ${\cal L}_G$ given by
Eq.(\ref{ehl}) and ${\cal L}_{\phi}$ by Eq.(\ref{lriem}). Variation in
relation to the metric tensor $g^{\mu \nu}$ yields the
gravitational field equation
\be
{\stackrel{\circ}{R}}{}_{\mu \nu} - \frac{1}{2} \; 
g_{\mu \nu} \; {\stackrel{\circ}{R}} = \frac{8 \pi G}{c^4} \,
{\cal T}_{\mu \nu} \; ,
\ee
where 
$$
{\cal T}_{\mu \nu} = - \frac{2}{\sqrt{-g}} \, \frac{\delta {\cal L}_{\phi}}
{\delta g^{\mu \nu}} 
$$
is the energy--momentum tensor of the scalar field. In the riemannian
description of gravitation, therefore, the energy--momentum tensor of any
matter field, as for example a scalar field, is able to produce curvature.

Now, we look for a minimal coupling prescription which brings the free
lagrangian (\ref{lfree}) to a lagrangian written in terms of the
teleparallel structure of spacetime. This prescription is given by
\ba
\eta^{a b} \; &\longrightarrow& \; \eta^{a b} \label{newmcp0} \\
{} \nonumber \\
\partial_a  \; &\longrightarrow& \; D_a = h_{a}{}^{\mu} \; D_\mu  \; , 
\label{newmcp}
\ea
where 
\be
D_\mu = \nabla_\mu - K_\mu
\label{dmu}
\ee
is the teleparallel version of the covariant derivative, with $\nabla_\mu$
the Cartan covariant derivative, and $K_\mu$ the contorsion tensor.
Therefore, in terms of the teleparallel structure, the scalar field
lagrangian assumes the form
\be
{\cal L}_{\phi} = \frac{h}{2} \left[ \eta^{a b} \, D_a \phi \;
D_b \phi - m^2 \phi^2 \right] \; ,
\label{lweitz}
\ee
where, for the specific case of a scalar field, 
\be
D_a = h_a{}^\mu \, \partial_\mu \; .
\label{da}
\ee
Using the identity
$$
\partial_\mu h = h h_a{}^{ \rho} \partial_\mu h^a{}_\rho
\equiv h \Gamma^{\rho}{}_{\rho \mu} \; ,
$$
it is easy to show that the corresponding field equation is
\be
{\Box} \phi + m^2 \phi = 0 \; ,
\label{klein1}
\ee
where
\be
{\Box} \phi = \left( \partial_\mu + {\Gamma}^{\rho }{}_
{\rho \mu} \right) \, \partial^\mu \phi \equiv h^{-1} \; 
\partial_\rho \left(h \; \partial^\rho \phi \right) 
\label{telb1}
\ee
is the teleparallel version of the Laplace--Beltrami operator. Because
${\Gamma}^{\rho }{}_{\rho \mu}$ is not symmetric in the last two
indices, $(\partial_\mu + {\Gamma}^{\rho }{}_{\rho \mu})$ is not the 
expression for the Cartan covariant divergence of $\partial^\mu \phi$.
By using Eq.(\ref{tor}), however, the expression for $\Box \phi$ may
be rewritten in the form
\be
\Box \phi = \left( \nabla_\mu + T^{\rho}{}_{\mu \rho} \right)
\partial^\mu \phi \; ,
\label{telb2}
\ee
where
$$
\nabla_\mu = \partial_\mu + {\Gamma}^{\rho }{}_{\mu \rho}
$$
is now the correct expression for the Cartan covariant divergence of
$\partial^\mu \phi$. Making use of the identity
\be
T^{\rho}{}_{\mu \rho} = - K^{\rho}{}_{\mu \rho} \; ,
\ee
easily obtained from Eq.(\ref{conto}), the teleparallel version of the
scalar field equation of motion is
\be
D_{\mu} \partial^\mu \phi + m^2 \phi = 0 \; ,
\label{klein3}
\ee
with $D_\mu$ the teleparallel covariant derivative (\ref{dmu}), here applied
to the spacetime vector field $\partial^\mu \phi$. Besides justifying the form
of the teleparallel minimal coupling prescription, therefore, we find that,
also in this case, it is completely equivalent to apply the minimal
coupling prescription in the lagrangian or in the field equation.

Now, Eq.(\ref{klein3}) can be rewritten as
\be
\nabla_\mu \, \partial^\mu \, \phi + m^2 \phi = - T^{\rho}{}_{\mu \rho} \, 
\partial^\mu \phi \equiv K^{\rho}{}_{\mu \rho} \, \partial^\mu \phi \; .
\label{klein2} 
\ee
In this form, it shows clearly that a scalar field, through its derivative
$\partial^\mu \phi$, couples to, and therefore feels torsion. Moreover, it
reveals that torsion plays a role similar to an external
force~\cite{paper}, quite analogous to the role played by the
electromagnetic field in the Lorentz force equation. On the other hand, from
Eqs.(\ref{gmn}) and ({\ref{car}) we have
$$
\Gamma_{\rho \lambda \mu} = - \Gamma_{\lambda \rho \mu} + \partial_\mu
g_{\rho \lambda} \; .
$$
Thus, in a locally inertial coordinate system, the Cartan connection
becomes skew symmetric in the first two indices, ${\Gamma}^{\rho }{}_{\rho
\mu}$ consequently vanishes, and the teleparallel version of the
Laplace--Beltrami operator becomes the free--field d'Alambertian operator.
This is the teleparallel version of the (strong) equivalence principle.

Let us consider again the total lagrangian ${\cal L}={\cal L}_G + {\cal
L}_{\phi}$, but now with ${\cal L}_G$ given by Eq.(\ref{lagr3}), and ${\cal
L}_{\phi}$ by Eq.(\ref{lweitz}). Variation in relation to the tetrad field
yields the teleparallel version of the gravitational field equation, which
can be written in the form~\cite{paper}
\be
\partial_\rho \; S_{\mu}{}^{\nu \rho} - 
\frac{4 \pi G}{c^4} \; t_{\mu}{}^{\nu} = \frac{4 \pi G}{c^4} \;
{\cal T}_{\mu}{}^{\nu} \; ,
\label{ym}
\ee
where $t_{\mu}{}^{\nu}$ is the energy--momentum (pseudo) tensor of the
gravitational field,
$$
{\cal T}_{\mu}{}^{\nu} = h^{\alpha}{}_{\mu} \left( - \frac{1}{h} \,
\frac{\delta {\cal L}_{\phi}}{\delta h^{\alpha}{}_{\nu}} \right)  
$$
is the energy--momentum tensor of the scalar field, and
$$
S_{\mu}{}^{\nu \rho} = \frac{1}{4} \; \left(T_{\mu}{}^{\nu \rho} +
T^{\nu}{}_{\mu}{}^{\rho} - T^{\rho}{}_{\mu}{}^{\nu}\right) - \frac{1}{2} \;
\left(\delta_{\mu}{}^{\rho} \; T_{\theta}{}^{\nu \theta} -
\delta_{\mu}{}^{\nu} \; T_{\theta}{}^{\rho \theta} \right) \; .
$$
In the teleparallel description of gravitation, therefore, energy and momentum
are the source of the dynamical torsion, a point which is not in
agreement with the usual belief that only a spin distribution could produce a
torsion field~\cite{ham}. Similar results have already been obtained in the
literature~\cite{saa}, being in fact the correct source of torsion still an open
problem. 

We remark once more that the Levi--Civita and the Cartan
connections are both defined on the very same spacetime, a manifold
endowed with both a riemannian and a teleparallel structures. Moreover, it
is possible to go from the riemannian to the teleparallel description
through very simple manipulations. For example, take the teleparallel
lagrangian (\ref{lweitz}), and substitute the covariant derivative
(\ref{da}). Then, by using Eq.(\ref{gmn}), one can easily see that this
lagrangian reduces to the riemannian lagrangian (\ref{lriem}). Obviously,
the same is true for the field equations: if we substitute  relation
(\ref{rel}) in the teleparallel Laplace--Beltrami (\ref{telb1}), it is an
easy task to verify that it reduces to the riemannian Laplace--Beltrami
(\ref{labe}). We notice furthermore that, in both descriptions, it results
completely equivalent to use the minimal coupling prescription in the
lagrangians or in the equations of motion.

\section{Final Remarks}

It has been known since long that only particles with spin could detect 
the teleparallel geometry, scalar matter being able to feel the metric
geometry only~\cite{nitsch}. However, as we have seen, the interaction
of gravitation with scalar matter can be described alternatively in
terms of magnitudes related to the riemannian or to the teleparallel
structures defined in spacetime, which are structures related respectively
to the Levi--Civita and the Cartan connections. Ultimately, this means that
scalar matter is able to feel anyone of these geometries. In other words,
scalar matter, through its derivative $\partial^\mu \phi$, is able to feel,
and therefore couples to torsion. Furthermore, based on the equivalence of
the corresponding lagrangians and field equations, we conclude that the
description in terms of the teleparallel geometry is completely equivalent
to the description in terms of the riemannian geometry. This means that,
besides coupling to torsion, the scalar field, trough its energy--momentum
tensor, can also be the source of torsion. It should be
remarked, however, that in the teleparallel description, the gravitational
interaction is not geometrized in the sense it is in general relativity,
presenting characteristics quite analogous to those provided by gauge
theories~\cite{hene}, with torsion playing the role of force. It is also
important to remark that, according to this approach, no Riemann--Cartan
geometry enters into the description of the gravitational interaction. As a
matter of fact, no experimental evidence seems to indicate the necessity of
including torsion, besides curvature, to correctly account for the gravitational
interaction .

Finally, it is worth mentioning that the minimal coupling prescription
(\ref{newmcp0})--(\ref{dmu}), introduced here to describe a gravitationally
coupled scalar field in the framework of the teleparallel equivalent of general
relativity, can be consistently applied to other fields as well. In the case of
the spin--one Maxwell field, for example, it is obtained as a consequence of its
application that, besides being able to be minimally coupled to torsion, the
electromagnetic field, through its energy--momentum tensor, can also produce
torsion. Furthermore, this coupling of the electromagnetic field with torsion is
found to preserve the local gauge invariance of Maxwell's theory, yielding in
this way a consistent description of such interaction~\cite{paper2}.

\section*{Acknowledgments}

The authors would like to thank R. Aldrovandi and L. C. B. Crispino for
useful discussions. They would also like to thank FAPESP--Brazil, and
CNPq--Brazil, for financial support.

\end{document}